Groupe d'Annecy

Laboratoire
d'Annecy-le-Vieux de
Physique des Particules

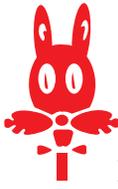 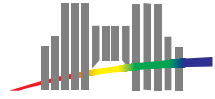

ENSLAPP

Groupe de Lyon

Ecole Normale
Supérieure de Lyon# The gravitational polarization tensor of thermal $\lambda\phi^4$ theory

Herbert Nachbagauer[*]
*Laboratoire de Physique Théorique ENSLAPP* [†]
*B.P. 110, F-74941 Annecy-le-Vieux Cedex, France*

Anton K. Rebhan[‡]
*DESY, Gruppe Theorie,*
*Notkestraße 85, D-22603 Hamburg, Germany*

Dominik J. Schwarz [§]
*Institut für Theoretische Physik, Technische Universität Wien,*
*Wiedner Hauptstraße 8-10/136, A-1040 Wien, Austria*
(July 19, 1995)## Abstract

The low-momentum structure of the gravitational polarization tensor of an ultrarelativistic plasma of scalar particles with $\lambda\phi^4$ interactions is evaluated in a two-loop calculation up to and including order $\lambda^{3/2}$. This turns out to require an improved perturbation theory which resums a local thermal mass term as well as nonlocal hard-thermal-loop vertices of scalar and gravitational fields.

[*]e-mail: herby@lapphp1.in2p3.fr

[†]URA 14-36 du CNRS, associée à l'E.N.S. de Lyon, et au L.A.P.P. (IN2P3-CNRS) d'Annecy-le-Vieux.

[‡]On leave of absence from Institut für Theoretische Physik der Technischen Universität Wien until November 1, 1995; e-mail: rebhana@x4u2.desy.de

[§]Address after October 1, 1995: Theoretische Physik, ETH–Hönggerberg, CH-8093 Zürich; e-mail: dschwarz@tph.tuwien.ac.athep-th/9507099  19 Jul 951

# I. INTRODUCTION

The central quantity in a linear response analysis of a thermal field theory is the polarization tensor and in particular its infrared behavior. It determines the spectrum of the quasi-particles and their dispersion laws, and also the dynamical and static screening of external fields. In the physics of the very early universe, which is filled with a hot plasma of various elementary particles, the gravitational polarization tensor is also of interest. It describes the response of the plasma to metric perturbations and its infrared behavior determines the dynamics of large-scale cosmological perturbations.

In Ref. [1], one of the present authors has calculated the leading temperature contributions to the gravitational polarization tensor $\Pi^{\alpha\beta\mu\nu}$ of a collisionless ultrarelativistic (i.e. effectively massless) plasma for temperatures $T \ll m_{Planck}$, where perturbation theory becomes applicable. Because the underlying effective action turns out to be conformally invariant, $\Pi^{\alpha\beta\mu\nu}$ can be calculated with flat-space momentum techniques and the corresponding tensor on a curved background space-time with vanishing Weyl tensor is obtained by a simple conformal transformation. In Ref. [2] this has been used to establish self-consistent equations for linear perturbations of Robertson-Walker cosmological models and exact analytic solutions have been constructed.

In the ultrarelativistic, collisionless case, the leading temperature contributions turn out to have a universal form, only the overall normalization changes in accordance with the magnitude of energy density. Some subleading contributions, where it starts to make a difference which particles constitute the plasma, have been obtained in Ref. [3].

In the following, we are interested in the corrections to the leading temperature terms when weak self-interactions are switched on. We take the simplest 4-dimensional model, scalar fields with $\lambda\phi^4$ interactions, which allows us to carry the calculations through resummed two-loop order. The emphasis of this paper is on the technical aspects of finite-temperature perturbation theory such as the need for resummation in order to extract the contributions proportional to $\lambda^{3/2}$. It will turn out that in the presence of gravitational interactions it is no longer sufficient to resum a local thermal mass term only. As in the Braaten-Pisarski resummation scheme for finite-temperature gauge theories, it becomes necessary to include nonlocal vertex corrections which are generated by "hard thermal loops" [4].

The implications of our results for self-consistent cosmological perturbations will be dealt with at length in a separate, forthcoming paper [5]; some first results have already appeared in Ref. [6].

# II. THE THERMAL GRAVITON POLARIZATION TENSOR

As thermal matter we take massless scalar particles with quartic self-interactions and interactions with the gravitational field according to the Lagrangian[1]

---

[1] Our conventions are those of Ref. [7]. In particular the metric signature is taken such that $g_{00} > 0$.

$$\mathcal{L}(x) = \sqrt{-g(x)} \left\{ \tfrac{1}{2} g^{\mu\nu} \partial_\mu \phi \partial_\nu \phi - \tfrac{1}{2} \xi R \phi^2 - \lambda \phi^4 \right\}. \tag{2.1}$$

If $\Gamma$ denotes all contributions to the effective action besides the classical Einstein-Hilbert action, the energy-momentum tensor is given by the one-point 1PI vertex function

$$T_{\mu\nu}(x) = \frac{2}{\sqrt{-g(x)}} \frac{\delta \Gamma}{\delta g^{\mu\nu}(x)} \tag{2.2}$$

and the gravitational polarization tensor by the two-point function

$$\Pi_{\mu\nu\alpha\beta}(x,y) \equiv \frac{\delta^2 \Gamma}{\delta g^{\mu\nu}(x) \delta g^{\alpha\beta}(y)} = \frac{1}{2} \frac{\delta \left( \sqrt{-g(x)} T_{\mu\nu}(x) \right)}{\delta g^{\alpha\beta}(y)}. \tag{2.3}$$

From the last equality it is clear that $\Pi^{\alpha\beta\mu\nu}$ describes the response of the (thermal) matter energy-momentum tensor to perturbations in the metric. Equating $\Pi^{\alpha\beta\mu\nu}$ to the perturbation of the Einstein tensor gives self-consistent equations for metric perturbations and, in particular, cosmological perturbations.

With massless scalars and $\xi = +1/6$, the matter part is conformally invariant, that is $\Gamma[g] = \Gamma[\Omega^2 g]$ — apart from the conformal anomaly which like other renormalization issues can be neglected in the high-temperature domain on which we shall concentrate.

This conformal invariance is crucial for our prospected applications for two reasons. Firstly, it allows us to have matter in thermal equilibrium despite a space-time dependent metric. As long as the latter is conformally flat, $ds^2 = \sigma(\tau, x)[d\tau^2 - dx^2]$, the local temperature on the curved background is determined by the scale factor $\sigma$. Secondly, the thermal correlation functions are simply given by the conformal transforms of their counterparts on a flat background, so that ordinary momentum-space techniques can be employed for their evaluation.

The energy-momentum tensor through two-loop order is given by the diagrams of Fig. 1 which are easily evaluated in flat space-time, e.g. within the imaginary-time formalism by choosing periodic boundary conditions for the scalar field in imaginary time with a period equal to the inverse temperature [8],

$$T_{\mu\nu}\Big|_{g=\eta} = \left[ \frac{\pi^2}{90} - \frac{\lambda}{48} + \ldots \right] T^4 \left( 4 \delta_\mu^0 \delta_\nu^0 - \eta_{\mu\nu} \right). \tag{2.4}$$

With $\eta \to g = \sigma \eta$ one has $T_{\mu\nu} \to \sigma^{-1} T_{\mu\nu}$, $\delta_\mu^0 \to \sigma^{1/2} \delta_\mu^0 = u_\mu$ as the normalized velocity field of the plasma, so that the temperature in curved space-time is obtained by $T \to \sigma^{-1/2} T$.

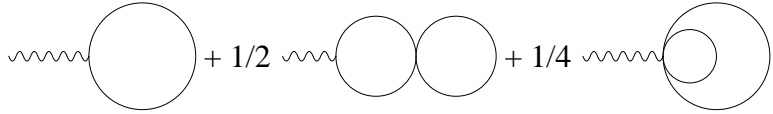

FIG. 1. One- and two-loop diagrams contributing to the energy-momentum tensor of $\lambda \phi^4$ theory. Wavy lines denote external gravitons and straight lines scalar particles.

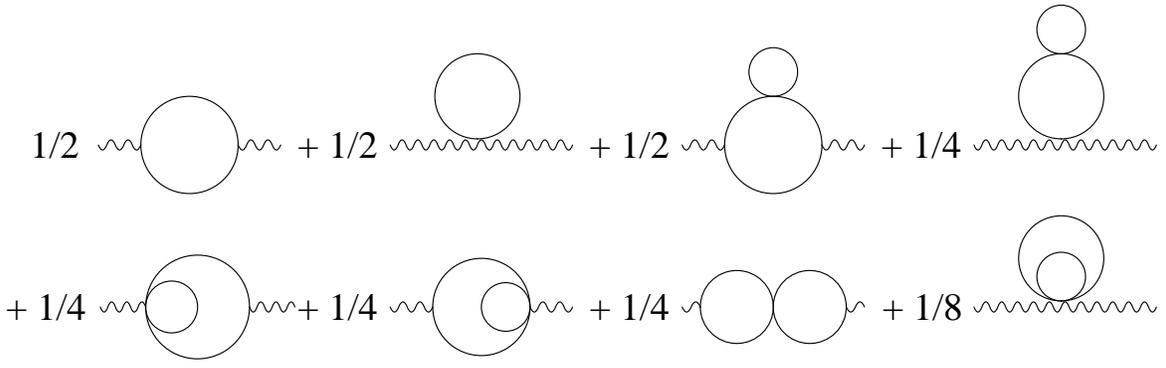

FIG. 2. The one- and two-loop diagrams of the gravitational polarization tensor.

The gravitational polarization tensor is an inherently nonlocal object, already in flat space-time, as is apparent from its diagrammatic expansion, shown in Fig. 2. On curved, but conformally flat, space it is then determined by

$$\Pi_{\mu\nu\alpha\beta}(x,y)\Big|_{g=\sigma\eta} = \sigma(x)\sigma(y)\Pi_{\mu\nu\alpha\beta}(x-y)\Big|_{g=\eta}. \tag{2.5}$$

Denoting by $\tilde{\Pi}_{\mu\nu\alpha\beta}(Q)$ the Fourier transform of $\Pi_{\mu\nu\alpha\beta}(x-y)\Big|_{g=\eta}$, the numerous components of the polarization tensor are restricted by diffeomorphism invariance through the Ward identity

$$4Q^\mu \tilde{\Pi}_{\mu\nu\alpha\beta}(Q) = Q_\nu T_{\alpha\beta} - Q^\sigma \left(T_{\alpha\sigma}\eta_{\beta\nu} + T_{\beta\sigma}\eta_{\alpha\nu}\right) \tag{2.6}$$

and by conformal invariance through the Weyl identity

$$\eta^{\mu\nu}\tilde{\Pi}_{\mu\nu\alpha\beta}(Q) = \frac{1}{2}T_{\alpha\beta}. \tag{2.7}$$

Using $\eta^{\mu\nu}$, $u_\mu = \delta^0_\mu$, and $Q^\mu = (Q^0, q)$, one can build 14 tensors to form a basis for $\tilde{\Pi}^{\mu\nu\alpha\beta}(Q) = \rho \sum_{i=1}^{14} c_i(Q) T_i^{\mu\nu\alpha\beta}(Q)$, see Table I. The above identities, however, reduce the number of independent structure functions to three. Choosing ($\rho = T_{00}$)

$$A(Q) \equiv \tilde{\Pi}_{0000}(Q)/\rho, \quad B(Q) \equiv \tilde{\Pi}_{0\mu}{}^\mu{}_0(Q)/\rho, \quad C(Q) \equiv \tilde{\Pi}_{\mu\nu}{}^{\mu\nu}(Q)/\rho \tag{2.8}$$

the $c_{1...14}$ are determined by the linear combinations given in Table II.

### III. ORDINARY PERTURBATIVE CALCULATION

A perturbative expansion in Feynman diagrams makes sense for temperatures $T \ll G^{-1/2} \sim m_{Planck}$. In this regime higher-order diagrams generated by additional graviton lines are correspondingly suppressed. Apart from this restriction, we shall concentrate on a high-temperature expansion in the sense of $T \gg Q_0, |q|(=q)$, since the typical length scale in the theory of cosmological perturbations is given by the horizon scale $\sim (GT^4)^{-1/2} \gg T^{-1}$.

The one-loop diagrams for the energy-momentum tensor and the gravitational polarization tensor describe collisionless matter. Their leading high-temperature contributions have been obtained first in Ref. [1] and the complete underlying effective action has been found subsequently in Ref. [9]. Here one may also include a thermal graviton background, since the leading-temperature contributions are proportional to the number of spin degrees of freedom, so gravitons are as important as any other thermalized matter. The one-loop result for the high-temperature gravitational polarization tensor is in fact determined by only one independent structure function, since it is constructed from a totally symmetric tensor

$$I_{\mu\nu\alpha\beta} = \oint_K \frac{K_\mu K_\nu K_\alpha K_\beta}{K^2(K-Q)^2}, \qquad \oint_K \equiv \sum_{K_0=2\pi i n T} \int \frac{d^3k}{(2\pi)^3},$$

according to

$$\tilde{\Pi}^{(1)}_{\mu\nu\alpha\beta}(Q) = \tfrac{1}{2} I_{\mu\nu\alpha\beta} - \tfrac{1}{4} \left( \delta^\sigma_\alpha \delta^\rho_\mu \eta_{\nu\beta} + \delta^\sigma_\alpha \delta^\rho_\nu \eta_{\mu\beta} + \delta^\sigma_\beta \delta^\rho_\mu \eta_{\nu\alpha} + \delta^\sigma_\beta \delta^\rho_\nu \eta_{\mu\alpha} \right) I^\gamma{}_{\gamma\sigma\rho}. \tag{3.1}$$

Therefore the functions $B$ and $C$ are directly related to the momentum-independent energy density, to wit,

$$B^{(1)} = -1, \qquad C^{(1)} = 0,$$

and all the nonlocalities reside in

$$A^{(1)}(Q) = \omega\, artanh \frac{1}{\omega} - \frac{5}{4} \tag{3.2}$$

with $\omega \equiv Q_0/q$.

This function has a logarithmic branch cut between $\omega = \pm 1$. The discontinuity along this cut is similar to the one found in the polarization tensor of gauge theories where it is well known to correspond to the phenomenon of Landau damping. Indeed, in the application of the above results to the theory of cosmological perturbations [2,10] it has been found that the discontinuity of $A$ is responsible for the collisionless damping of subhorizon perturbations.

At two-loop order one begins to see the effects of self-interactions of the thermal matter. If $\lambda \gg \sqrt{G}T \sim T/m_{Planck}$, which we shall assume, only the matter self-interactions are important. A straightforward (albeit somewhat tedious) evaluation of the diagrams of Fig. 2 yields

$$A^{(2)}(Q) = \frac{5\lambda}{8\pi^2} \left[ 2 \left( \omega\, artanh \frac{1}{\omega} \right)^2 - \omega\, artanh \frac{1}{\omega} - \frac{\omega^2}{\omega^2-1} \right], \tag{3.3}$$

$$B^{(2)}(Q) = \frac{5\lambda}{4\pi^2} \left[ -(\omega^2-1)\left( \omega\, artanh \frac{1}{\omega} \right)^2 + (2\omega^2-1)\omega\, artanh \frac{1}{\omega} - \omega^2 \right], \tag{3.4}$$

$$C^{(2)}(Q) = \frac{5\lambda}{8\pi^2} \left[ 3(\omega^2-1)^2 \left( \omega\, artanh \frac{1}{\omega} \right)^2 - 2(\omega^2-1)(3\omega^2-2)\omega\, artanh \frac{1}{\omega} + 3\omega^4 - 4\omega^2 \right]. \tag{3.5}$$

In addition to a logarithmic branch cut between $\omega = \pm 1$, whose discontinuity is now more complicated, there are also simple poles in $A$ at $\omega = \pm 1$.

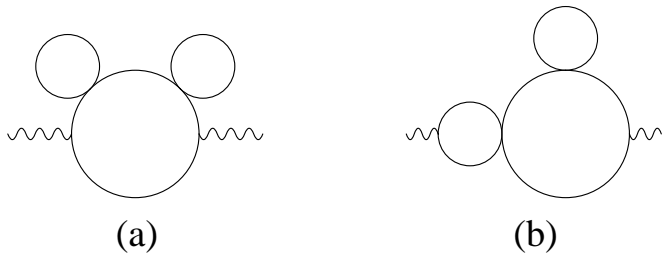

FIG. 3. Two examples of infrared divergent graphs beyond two-loop order.

## IV. RESUMMATION

If one now attempts to go further, one encounters a problem with ordinary perturbation theory starting at three-loop order. Some of the diagrams, e.g. those shown in Fig. 3, are infrared divergent.

The scalar self-energy subdiagrams in Fig. 3a are proportional to $\lambda T^2$ and their repeated insertion generates a chain of massless propagators all with the same momentum. This type of divergence and its treatment are well-known [11]. The interactions of the (massless) scalar particles with the plasma generate a thermal mass which needs to be resummed. However, in addition to repeated insertion of thermal masses there are also 1PI vertex subdiagrams which can cause infrared divergences. In Fig. 3b the vertex correction has a similar effect as the second mass insertion in Fig. 3a — both make the respective diagram infrared divergent.

It is therefore necessary to include the (generally nonlocal) vertex corrections when curing perturbation theory by a resummation of thermal masses. This can also be understand by noting that a simple local mass term would break conformal invariance. But apart from the zero-temperature conformal anomaly the full energy-momentum tensor has still to be traceless for conformally coupled scalars, which shows that thermally induced masses are not equivalent to ordinary rest masses, not even for scalar fields where the thermal mass is independent of momentum.

A systematic way to improve perturbation theory when vertex corrections are of equal importance has been worked out first for high-temperature quantum chromodynamics by Braaten and Pisarski [4]. It has been found there that it is necessary to resum all the subdiagrams which are generated by loop momenta $\sim T$ (termed "hard thermal loops"). In $\lambda\phi^4$ theory without gravitational interactions, the only hard thermal loop is given by the scalar self-energy diagram, Fig. 4a, which yields a constant mass term $m^2 = \lambda T^2$. Clearly, when probing momentum scales $\lesssim \sqrt{\lambda}T$, this has to be resummed. In the presence of external gravitational fields, vertex diagrams like the ones in Fig. 4b are of comparable magnitude, however they turn out to be nonlocal.

In Ref. [12] it has been shown how to construct the entire generating functional $\Delta S[\phi, g]$ for hard thermal loops in the presence of gravity. In the case of $\lambda\phi^4$ theory this becomes particularly simple [13]. The hard thermal loop effective action can be written as an integral over a forward scattering amplitude $J[K; \phi, g]$ according to

$$\Delta S[\phi,g] = \frac{1}{(2\pi)^3} \int d^4K \, \delta(K^2)\theta(K_0)n(K_0)J[K;\phi,g]. \tag{4.1}$$

where $n(K_0) = (\exp(K_0/T) - 1)^{-1}$ denotes the Bose-Einstein factor. It then turns out that all the dependence of $J$ on the metric can be eliminated by choosing the diffeomorphism gauge condition

$$K^\mu \hat{g}_{\mu\nu} = K^\mu \eta_{\mu\nu}, \tag{4.2}$$

whereupon

$$J[K;\phi,\hat{g}] = -6\lambda \int d^4x \, \phi^2(x). \tag{4.3}$$

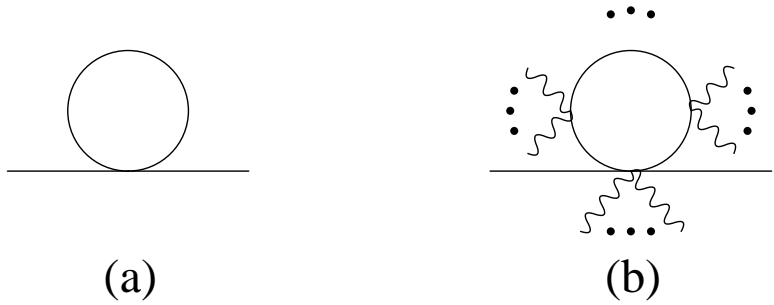

FIG. 4. (a) The scalar self-energy which gives rise to the thermal mass. (b) The general form of the non-local vertex diagrams which have to be resummed when going beyond $O(\lambda)$.

Assuming an asymptotically conformally flat space, one can reconstruct the generally covariant function J in terms of the set of null geodesics $y^\mu(x,\theta)$, $y^\mu(x,0) = x^\mu$, whose tangent vector $\dot{y}^\mu(x,\theta) \to K^\mu$ for affine parameter $\theta \to -\infty$. The solution is

$$\Delta S[\phi,g] = -\frac{1}{2}\lambda T^2 \int \frac{d\Omega_K}{4\pi} \left[ \int d^4x \sqrt{-g(x)} \phi^2(x) e^{-\int_{-\infty}^0 d\theta \nabla_\mu \dot{y}^\mu(x,\theta)} \right]. \tag{4.4}$$

The integral over the geodesic divergence is clearly a highly nonlocal quantity and is what reconciles the appearance of a thermal mass for the scalars with invariance under Weyl rescalings of the metric.

The leading high-temperature contributions to the graviton self-energy and higher vertex functions with only external graviton lines can also be summarized in a similar (but more complicated) generating functional [12]. However, we shall not need it in our resummation program since we assumed $\lambda \gg (T/m_{Planck})^2$, by which we can neglect higher-order diagrams with internal graviton lines.

A resummation of all the hard thermal loops summarized by $\Delta S[\phi,g]$ is formally achieved by writing $\int d^4x \mathcal{L} \equiv S = S_{res.} - S_{counter}$, with $S_{res.} = S + \Delta S$ and $S_{counter} = \Delta S$. The Feynman rules now include the thermal mass of the scalars and the additional hard-thermal-loop vertices which have two scalar lines and an arbitrary number of graviton lines. Overcounting is avoided by subtracting them as counterterms, as which they are treated as one-loop objects. [Equivalently, one could consider $\Delta S$ as coming from integrating out all the modes with hard momenta $\gtrsim T$ after which $S + \Delta S$ is the effective action to be taken when integrating out the remaining soft modes.]

Performing a resummed two-loop calculation, we can collect all terms up to but excluding order $\lambda^2$, which is the formal order of bare three-loop diagrams. Since the resummation introduces $\sqrt{\lambda}$ through $m = \sqrt{\lambda}T$, we can go up to order $\lambda^{3/2}$, where the plasmon effect enters for the first time. [Because the bare two-loop diagrams are infrared convergent, there are no contributions of order $\lambda^{1/2}$.] At two-loop order one has to include the thermal counterterms. Here it turns out that the diagrammatics is considerably simplified if the hard-thermal-loop mass $\sqrt{\lambda}T$ in $\Delta S$ is replaced by its resummed value, which is given by the gap equation

$$m^2 = -12\lambda \sumint_K \frac{1}{K^2 - m^2} = \lambda T^2 - \frac{3}{\pi}\lambda^{3/2}T^2 + \ldots. \qquad (4.5)$$

Then thermal counterterms can cancel entire subdiagrams, while the sum of the resummed two-loop diagrams is unchanged up to order $\lambda^2$.

### A. The energy-momentum tensor up to order $\lambda^{3/2}$

When switching to the resummed Feynman rules, the one- and two-loop diagrams for the energy momentum tensor of Fig. 1 are replaced by the ones shown in Fig. 5a-c, where a dot indicates a dressed propagator or a dressed vertex, respectively. In addition there are two one-loop diagrams involving thermal counterterms (denoted by crosses), 5d and 5e. The mass counterterm in diagram 5e cancels diagram 5b completely (using Eq. (4.5)), bringing down the total number of diagrams to two. The vertex counterterm in diagram 5d reduces the full vertex in diagram 5a to a bare vertex, so that the resummation of the vertices turns out to be superfluous in this case, for there is no hard-thermal-loop vertex which could modify the vertex in diagram 5c.

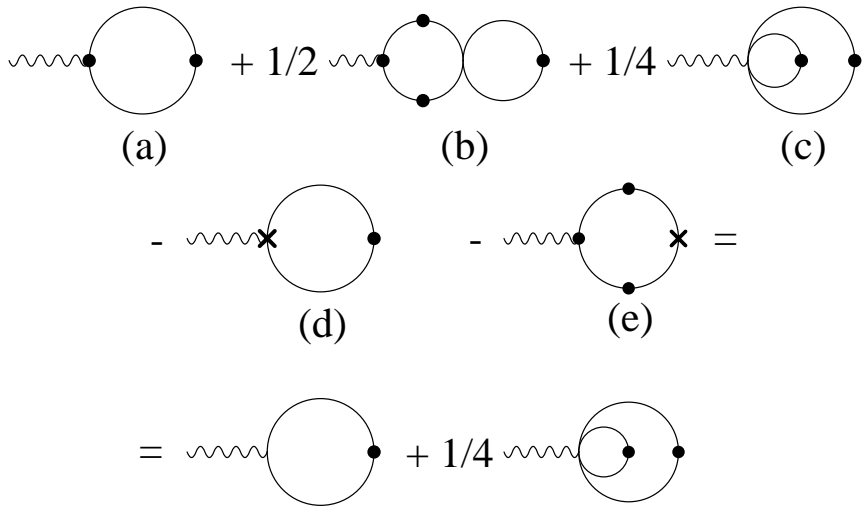

FIG. 5. The resummed energy-momentum tensor. The dots denote the dressed propagators and the dressed vertices, whereas the crosses indicate the thermal mass and vertex counterterms.

The net result thus reproduces the one obtained by simple ring resummation [8] and reads

$$T_{\mu\nu}\Big|_{g=\eta} = \oint_K \frac{\eta_{\mu\nu}K^2/2 - K_\mu K_\nu}{K^2 - m^2} + 3\lambda\eta_{\mu\nu}\left(\oint_K \frac{1}{K^2 - m^2}\right)^2$$
$$= \left[\frac{\pi^2}{90} - \frac{\lambda}{48} + \frac{\lambda^{3/2}}{12\pi} + O(\lambda^2)\right]T^4\left(4\delta^0_\mu\delta^0_\nu - \eta_{\mu\nu}\right), \quad (4.6)$$

which is traceless as it should be.

### B. The gravitational polarization tensor up to order $\lambda^{3/2}$

In contrast to the energy-momentum tensor, the resummation of the nonlocal hard-thermal-loop vertices turns out to be crucial to obtain the complete result for the gravitational polarization tensor up to and including order $\lambda^{3/2}$. A naive ring resummation of a local mass term for the scalars would lead to a violation of the Weyl identity (2.7).

The diagrammatics for the two-point function at resummed two-loop order is given in Fig. 6. As before, the two-loop diagrams which correspond to mass insertions can be canceled by thermal counterterms when using Eq. (4.5). However, this time the hard-thermal vertices do not cancel out completely, except the one in diagram 6b. In diagram 6a, there is only a partial cancellation with the vertex counterterms, and in the two-loop diagrams 6e-g the dressed vertices are untouched at this order.

In the application to the theory of cosmological perturbations, we shall need in particular the low-momentum limit $Q_0, q \sim \sqrt{G}T^2 \ll \sqrt{\lambda}T = m$. In this limit the results become independent of the coupling constant $\xi$ and read explicitly

$$\tilde{\Pi}_{\alpha\beta\gamma\delta}(Q) =$$
$$= -\frac{1}{8}\oint_K \Delta_K\Delta_P\left(K\cdot P\eta_{\alpha\beta} - K_\alpha P_\beta - P_\alpha K_\beta\right)\left(K\cdot P\eta_{\gamma\delta} - K_\gamma P_\delta - P_\gamma K_\delta\right)$$
$$-\frac{1}{8}\oint_K \Delta_K K^2\left(\eta_{\alpha\gamma}\eta_{\beta\delta} + \eta_{\alpha\delta}\eta_{\beta\gamma} - \eta_{\alpha\beta}\eta_{\gamma\delta}\right)$$
$$-\frac{1}{4}\oint_K \Delta_K\left[\eta_{\alpha\beta}K_\gamma K_\delta + \eta_{\gamma\delta}K_\alpha K_\beta - \eta_{\alpha\gamma}K_\beta K_\delta - \eta_{\alpha\delta}K_\beta K_\gamma - \eta_{\beta\gamma}K_\alpha K_\delta - \eta_{\beta\delta}K_\alpha K_\gamma\right]$$
$$+\frac{3}{2}\lambda\oint_K \Delta_K\Delta_P\left(K\cdot P\eta_{\alpha\beta} - K_\alpha P_\beta - P_\alpha K_\beta\right)\oint_K \Delta_K\Delta_P\left(K\cdot P\eta_{\gamma\delta} - K_\gamma P_\delta - P_\gamma K_\delta\right)$$
$$-\frac{3}{4}\lambda\left(\eta_{\alpha\gamma}\eta_{\beta\delta} + \eta_{\alpha\delta}\eta_{\beta\gamma} - \eta_{\alpha\beta}\eta_{\gamma\delta}\right)\left(\oint_K \Delta_K\right)^2$$
$$-\frac{3}{2}\lambda\eta_{\alpha\beta}\left(\oint_K \Delta_K\right)\oint_K \Delta_K\Delta_P\left(K\cdot P\eta_{\gamma\delta} - K_\gamma P_\delta - P_\gamma K_\delta\right)$$
$$-\frac{3}{2}\lambda\eta_{\gamma\delta}\left(\oint_K \Delta_K\right)\oint_K \Delta_K\Delta_P\left(K\cdot P\eta_{\alpha\beta} - K_\alpha P_\beta - P_\alpha K_\beta\right)$$
$$-\frac{1}{8}\eta_{\alpha\beta}\eta_{\gamma\delta}m^4\oint_K \Delta_K\Delta_P$$
$$-\frac{3}{2}\lambda m^2\eta_{\alpha\beta}\left(\oint_K \Delta_K\Delta_P\right)\oint_K \Delta_K\Delta_P\left(K\cdot P\eta_{\gamma\delta} - K_\gamma P_\delta - P_\gamma K_\delta\right)$$

$$-\frac{3}{2}\lambda m^2 \eta_{\gamma\delta} \left(\oint_K \Delta_K \Delta_P\right) \oint_K \Delta_K \Delta_P \left(K \cdot P \eta_{\alpha\beta} - K_\alpha P_\beta - P_\alpha K_\beta\right)$$

$$+\frac{3}{2}\lambda m^4 \eta_{\alpha\beta}\eta_{\gamma\delta} \left(\oint_K \Delta_K \Delta_P\right)^2$$

$$-18\lambda^2 \left(\oint_K \Delta_K \Delta_P\right) \left(\oint_K \Delta_K \Delta_P \left(K \cdot P \eta_{\alpha\beta} - K_\alpha P_\beta - P_\alpha K_\beta\right)\right)$$

$$\times \oint_K \Delta_K \Delta_P \left(K \cdot P \eta_{\gamma\delta} - K_\gamma P_\delta - P_\gamma K_\delta\right), \tag{4.7}$$

where $\Delta_K = 1/(K^2 - m^2)$ and $P \equiv K - Q$.

With this expression one can explicitly verify that the Ward and Weyl identities are fulfilled. As a consequence the tensorial structure is as given in Table II, with three independent structure functions. Evaluating the integrals involved as described in the Appendix leads to our final result

$$A = \omega \, \mathrm{artanh}\frac{1}{\omega} - \frac{5}{4}$$
$$+ \frac{5\lambda}{8\pi^2}\left[2\left(\omega\,\mathrm{artanh}\frac{1}{\omega}\right)^2 - \omega\,\mathrm{artanh}\frac{1}{\omega} - \frac{\omega^2}{\omega^2-1}\right]$$
$$+ \frac{5\lambda^{3/2}}{8\pi^3}\left[3\left(\omega^2-1-\omega\sqrt{\omega^2-1}\right)\left(\omega\,\mathrm{artanh}\frac{1}{\omega}\right)^2\right.$$
$$+6\left(\omega\sqrt{\omega^2-1}-\omega^2-\frac{\omega}{\sqrt{\omega^2-1}}\right)\omega\,\mathrm{artanh}\frac{1}{\omega}$$
$$\left.+\frac{\omega}{(\omega^2-1)^{3/2}}+3\frac{\omega^2}{\omega^2-1}+6\frac{\omega}{\sqrt{\omega^2-1}}-3\omega\sqrt{\omega^2-1}+3\omega^2\right] \tag{4.8}$$

$$B = -1 + \frac{5\lambda}{4\pi^2}\left[-(\omega^2-1)\left(\omega\,\mathrm{artanh}\frac{1}{\omega}\right)^2 + (2\omega^2-1)\omega\,\mathrm{artanh}\frac{1}{\omega} - \omega^2\right]$$
$$+ \frac{15\lambda^{3/2}}{8\pi^3}\left[\left\{\omega(\omega^2-1)^{3/2} - (\omega^2-1)^2\right\}\left(\omega\,\mathrm{artanh}\frac{1}{\omega}\right)^2\right.$$
$$+2\left\{(\omega^2-1)^2 - \omega(\omega^2-1)^{3/2} + \omega\sqrt{\omega^2-1} - \omega^2\right\}\omega\,\mathrm{artanh}\frac{1}{\omega}$$
$$\left.-\frac{\omega}{\sqrt{\omega^2-1}} - 2\omega\sqrt{\omega^2-1} + \omega(\omega^2-1)^{3/2} - \omega^4 + 4\omega^2\right] \tag{4.9}$$

$$C = \frac{5\lambda}{8\pi^2}\left[3(\omega^2-1)^2\left(\omega\,\mathrm{artanh}\frac{1}{\omega}\right)^2 - 2(\omega^2-1)(3\omega^2-2)\omega\,\mathrm{artanh}\frac{1}{\omega} + 3\omega^4 - 4\omega^2\right]$$
$$+ \frac{15\lambda^{3/2}}{16\pi^3}\left[3(\omega^2-1)^2\left(\omega^2-1-\omega\sqrt{\omega^2-1}\right)\left(\omega\,\mathrm{artanh}\frac{1}{\omega}\right)^2\right.$$
$$-2(\omega^2-1)\left\{3(\omega^2-1)^2 - 3\omega(\omega^2-1)^{3/2} + 3\omega\sqrt{\omega^2-1} - 3\omega^2 + 1\right\}\omega\,\mathrm{artanh}\frac{1}{\omega}$$
$$\left.-3\omega(\omega^2-1)^{5/2} + 6\omega(\omega^2-1)^{3/2} + \omega\sqrt{\omega^2-1} + 3\omega^6 - 15\omega^4 + 16\omega^2\right]. \tag{4.10}$$

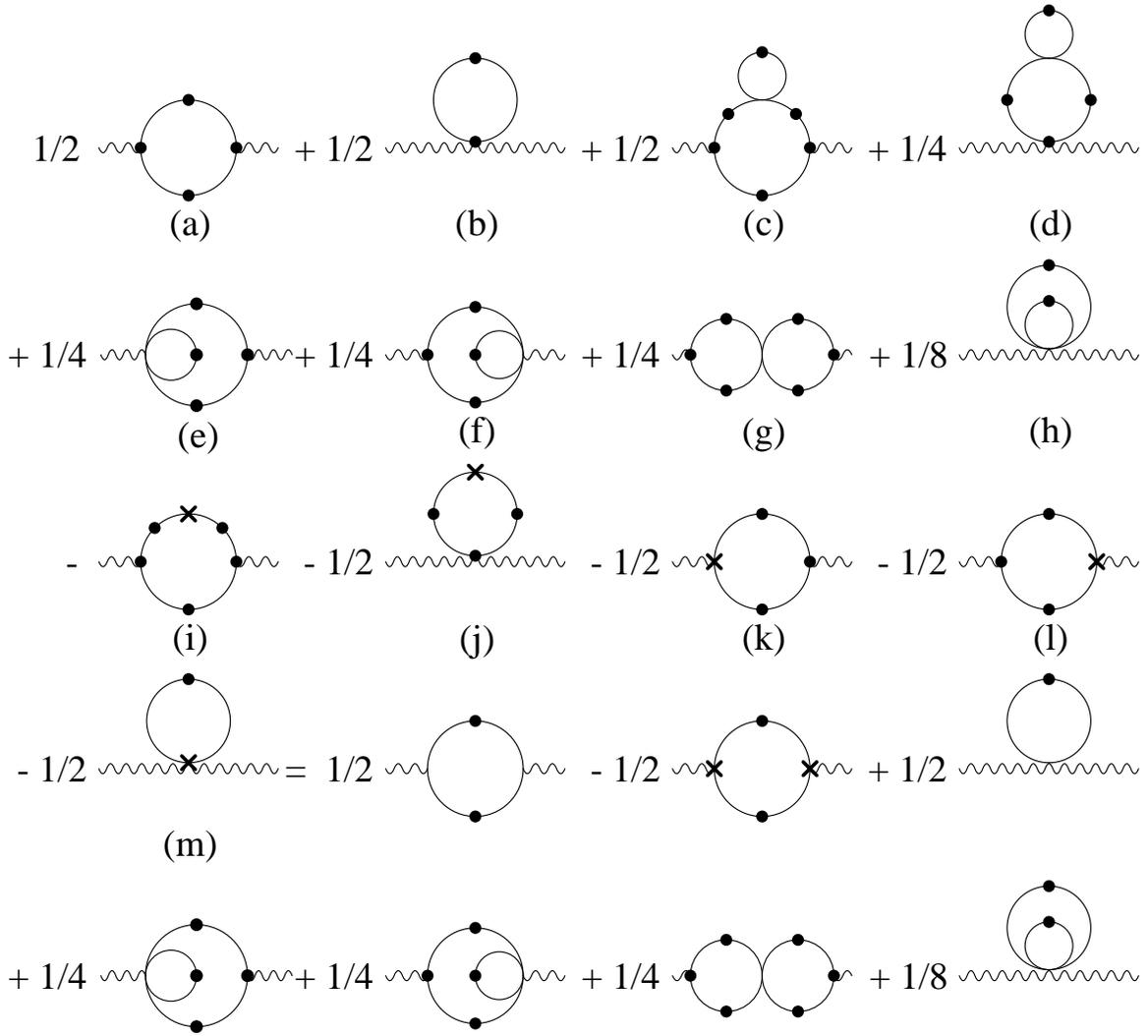

FIG. 6. The resummed gravitational polarization tensor.

A new feature of the contributions proportional to $\lambda^{3/2}$ is that in addition to poles at $\omega = \pm 1$ and a logarithmic branch cut in between there are now terms involving root singularities and a corresponding branch cut. The singularities themselves are in fact spurious, since the perturbative expansion breaks down for $|\omega| - 1 \lesssim \lambda$. As found in a somewhat different context in Ref. [14], the singularities at $\omega = \pm 1$ are removed by the thermal masses of the scalar particles, although the branch cut for $|\omega| < 1$ persists.

Consider the first diagram on the right-hand side of Fig. 6 and its contribution to the discontinuity of $A = \tilde{\Pi}_{0000}/\rho$,

$$\mathrm{Disc}\, A_1 = -\theta(1-\omega^2)\frac{\omega}{8\pi\rho}\int_{m/\sqrt{1-\omega^2}}^{\infty} dp\, n'(p)\left(p^2 - m^2/2\right)^2, \qquad (4.11)$$

where we have taken the limit $Q_0, q \ll m = \sqrt{\lambda}T$ but refrained from any further expansion.

The expansion in $\lambda$ as performed above gives

$$\mathrm{D}isc\, A_1 = \theta(1-\omega^2)\omega \left\{ \left(1 + \frac{5\lambda}{8\pi^2} - \frac{15\lambda^{3/2}}{4\pi^3}\right)\pi \right.$$
$$\left. + \frac{5\lambda^{3/2}}{4\pi^3}\left(\frac{3}{4}\sqrt{1-\omega^2} + \frac{3}{\sqrt{1-\omega^2}} - \frac{1}{(1-\omega^2)^{3/2}}\right) + O(\lambda^2) \right\}, \quad (4.12)$$

which reproduces the discontinuity of $A$ at leading order as well as that of the most singular term that ocured at order $\lambda^{3/2}$.

In the limit $1-\omega^2 \ll \lambda$, on the other hand, one can read off from Eq. (4.11) that

$$\mathrm{D}isc\, A_1 \to \theta(1-\omega^2)\frac{15\omega}{4\pi^3}\frac{\lambda^2}{(1-\omega^2)^2}\exp\left(-\sqrt{\frac{\lambda}{1-\omega^2}}\right), \quad (4.13)$$

so the discontinuous and even singular onset of the imaginary part of $A$ that appeared in Eq. (4.12) is illusory. It actually sets in smoothly, but nonanalytically, with all derivatives vanishing. Some consequences of this in the context of cosmological perturbations have been discussed in Ref. [6].

A more complete application of all the above results to the determination of the dynamics of cosmological perturbations in the presence of a weakly interacting ultrarelativistic plasma will be the subject of a forthcoming publication [5].

## ACKNOWLEDGMENTS

We are grateful to Eamonn Gaffney for useful discussions. This work was supported by the Austrian "Fonds zur Förderung der wissenschaftlichen Forschung (FWF)" under projects no. P9005-PHY and P10063-PHY, and by the EEC Programme "Human Capital and Mobility", contract CHRX-CT93-0357 (DG 12 COMA).

## APPENDIX: EVALUATION OF THE THERMAL INTEGRALS

The evaluation of the massive thermal integrals in Eqs. (4.8–4.10) was performed by a MATHEMATICA program. The principal steps are the following.

First a partial fraction decomposition in $k^0$ is performed, which gives a sum of terms in which $k^0$ appears only linearly in the denominators. After a shift $k^0 \to k^0 + q^0$ where necessary, the thermal sums are carried out with help of the formula (dropping zero temperature contributions)

$$T\sum_{k^0=2\pi i nT}\frac{1}{k^0 \pm E} = \pm n(E) \quad (A1)$$

The sum does not converge as it stands but it occurs only in combinations where the divergent part cancels and the above formula applies. The energy can take on the values $E_k = \sqrt{k^2 + m^2}$ or $\sqrt{(k-q)^2 + m^2}$ according to the loop momentum. In the latter case it is convenient to perform a shift of the integration variable $k \to k+q$ such that $E_k$ becomes the argument in the distribution function in all summands.

The spatial loop-integration is split into an integration over $k$ and the angle $kq/(kq)$ which is performed first. This leads to the following three types of integrals

$$\int_0^\infty \frac{dk\, k^l}{E_k}\, n(E_k), \quad l = 0,\ldots,4\,, \qquad (A2)$$

$$\int_0^\infty \frac{dk\, k^l}{E_k}\, n(E_k)\, \log\frac{(2k-q)^2 + \Omega^2}{(2k+q)^2 + \Omega^2}, \quad l = 1,3,5\,, \qquad (A3)$$

$$\int_0^\infty dk\, k^l\, n(E_k)\, \mathrm{artanh}\frac{8k\omega E_k}{4k^2(1+\omega^2) + 4m^2\omega^2 - q^2(\omega^2-1)^2}, \quad l = 1,3\,. \qquad (A4)$$

where for the moment we specified to the kinematical region $\Omega^2 = \omega^2(4m^2 - q^2(\omega^2-1))/(\omega^2-1) \leq 0$. The corresponding formulas in other regions can be obtained by analytic continuation of the final expressions.

The high temperature expansion of integrals of the form (A2) is well known [8,15] and will not be repeated here. The strategy for the remaining integrals is to subtract off the large momentum behaviour of the integrand until the subtracted integrand decreases sufficiently for large $k$ such that the Bose-Einstein distribution function can be replaced by $T/E_k$. For the integrals (A3), a suitable subtraction is given by the Taylor-series polynomial in $1/k$ of order $l$ of the logarithm.

On the other hand, for integrals of typ (A4) the subtractions have been chosen in a different way anticipating the subsequent integration. For $l=1$ it suffices to subtract the $k$-independent large momentum limit $2\,\mathrm{artanh}\,\omega^{-1}$ whereas in the case $l=3$ an additional second order subtraction proportional to $1/E_k^2$ was performed. The subtractions themselves give rise to integrals of typ (A2) without the energy in the denominator and may be evaluated along similar lines as the integrals (A2). The subtracted integral with the replaced Bose-Einstein distribution can be evaluated elementarily by introducing $E_k$ as integration variable.

It still remains to evaluate the order $T$ contribution of the typ (A3) integrals. For this purpose we take advantage of the fact that the integrand is even and integrate from $-\infty$ to $\infty$. Closing the integration contour in the upper half-plane $\Im m(k) \geq 0$ only picks up a cut contribution from the logarithm and a pole contribution at $k = im$ since on the arc the subtracted integrand vanishes in the limit $|k| \to \infty$.

The final expressions are obtained by replacing the thermal mass by its value given in (4.5) and expanding in powers of $\lambda^{1/2}$.

TABLES

$T_1^{\alpha\beta\mu\nu} = \eta^{\alpha\nu}\,\eta^{\beta\mu} + \eta^{\alpha\mu}\,\eta^{\beta\nu}$

$T_2^{\alpha\beta\mu\nu} = u^\mu\,\left(u^\beta\,\eta^{\alpha\nu} + u^\alpha\,\eta^{\beta\nu}\right) + u^\nu\,\left(u^\beta\,\eta^{\alpha\mu} + u^\alpha\,\eta^{\beta\mu}\right)$

$T_3^{\alpha\beta\mu\nu} = u^\alpha\,u^\beta\,u^\mu\,u^\nu$

$T_4^{\alpha\beta\mu\nu} = \eta^{\alpha\beta}\,\eta^{\mu\nu}$

$T_5^{\alpha\beta\mu\nu} = u^\mu\,u^\nu\,\eta^{\alpha\beta} + u^\alpha\,u^\beta\,\eta^{\mu\nu}$

$T_6^{\alpha\beta\mu\nu} = u^\beta\,\left(\bar{Q}^\nu\,\eta^{\alpha\mu} + \bar{Q}^\mu\,\eta^{\alpha\nu}\right) + \bar{Q}^\beta\,\left(u^\nu\,\eta^{\alpha\mu} + u^\mu\,\eta^{\alpha\nu}\right)$
$\qquad\qquad + u^\alpha\,\left(\bar{Q}^\nu\,\eta^{\beta\mu} + \bar{Q}^\mu\,\eta^{\beta\nu}\right) + \bar{Q}^\alpha\,\left(u^\nu\,\eta^{\beta\mu} + u^\mu\,\eta^{\beta\nu}\right)$

$T_7^{\alpha\beta\mu\nu} = \bar{Q}^\nu\,u^\alpha\,u^\beta\,u^\mu + \bar{Q}^\mu\,u^\alpha\,u^\beta\,u^\nu + \bar{Q}^\beta\,u^\alpha\,u^\mu\,u^\nu + \bar{Q}^\alpha\,u^\beta\,u^\mu\,u^\nu$

$T_8^{\alpha\beta\mu\nu} = \bar{Q}^\beta\,\bar{Q}^\nu\,\eta^{\alpha\mu} + \bar{Q}^\beta\,\bar{Q}^\mu\,\eta^{\alpha\nu} + \bar{Q}^\alpha\,\bar{Q}^\nu\,\eta^{\beta\mu} + \bar{Q}^\alpha\,\bar{Q}^\mu\,\eta^{\beta\nu}$

$T_9^{\alpha\beta\mu\nu} = \bar{Q}^\mu\,\bar{Q}^\nu\,u^\alpha\,u^\beta + \bar{Q}^\alpha\,\bar{Q}^\beta\,u^\mu\,u^\nu$

$T_{10}^{\alpha\beta\mu\nu} = \left(\bar{Q}^\beta\,u^\alpha + \bar{Q}^\alpha\,u^\beta\right)\,\left(\bar{Q}^\nu\,u^\mu + \bar{Q}^\mu\,u^\nu\right)$

$T_{11}^{\alpha\beta\mu\nu} = \bar{Q}^\beta\,\bar{Q}^\mu\,\bar{Q}^\nu\,u^\alpha + \bar{Q}^\alpha\,\bar{Q}^\mu\,\bar{Q}^\nu\,u^\beta + \bar{Q}^\alpha\,\bar{Q}^\beta\,\bar{Q}^\nu\,u^\mu + \bar{Q}^\alpha\,\bar{Q}^\beta\,\bar{Q}^\mu\,u^\nu$

$T_{12}^{\alpha\beta\mu\nu} = \bar{Q}^\alpha\,\bar{Q}^\beta\,\bar{Q}^\mu\,\bar{Q}^\nu$

$T_{13}^{\alpha\beta\mu\nu} = \bar{Q}^\mu\,\bar{Q}^\nu\,\eta^{\alpha\beta} + \bar{Q}^\alpha\,\bar{Q}^\beta\,\eta^{\mu\nu}$

$T_{14}^{\alpha\beta\mu\nu} = \left(\bar{Q}^\nu\,u^\mu + \bar{Q}^\mu\,u^\nu\right)\,\eta^{\alpha\beta} + \left(\bar{Q}^\beta\,u^\alpha + \bar{Q}^\alpha\,u^\beta\right)\,\eta^{\mu\nu}$

TABLE I. A basis of 14 independent tensors $T_i^{\alpha\beta\mu\nu}$ built from $\eta^{\mu\nu}$, $u^\mu = \delta_0^\mu$, and $\bar{Q}^\mu \equiv Q^\mu/q = (\omega, \mathbf{q}/q)$.

$c_1 = \frac{1}{8}\bar{Q}^4 A + \frac{1}{2}\bar{Q}^2 B + \frac{1}{4}C + \frac{1}{32}\bar{Q}^4 + \frac{11}{24}\bar{Q}^2 + \frac{1}{6}$

$c_2 = \frac{5}{8}\bar{Q}^6 A + \bar{Q}^4 B + \frac{1}{4}\bar{Q}^2 C + \frac{5}{32}\bar{Q}^6 + \frac{19}{24}\bar{Q}^4 + \frac{1}{12}\bar{Q}^2 - \frac{1}{3}$

$c_3 = \bar{Q}^2\,\{\frac{35}{8}\bar{Q}^6 A + \frac{5}{2}\bar{Q}^4 B + \frac{1}{4}\bar{Q}^2 C + \frac{35}{32}\bar{Q}^6 + \frac{25}{24}\bar{Q}^4 + \frac{7}{12}\bar{Q}^2 - \frac{1}{3}\}$

$c_4 = \bar{Q}^2\,\{\frac{1}{8}\bar{Q}^2 A - \frac{1}{2}B - \frac{1}{4}C + \frac{1}{32}\bar{Q}^2 - \frac{13}{24}\}$

$c_5 = \bar{Q}^2\,\{\frac{5}{8}\bar{Q}^4 A - \frac{1}{2}\bar{Q}^2 B - \frac{1}{4}C + \frac{5}{32}\bar{Q}^4 - \frac{17}{24}\bar{Q}^2 + \frac{1}{12}\}$

$c_6 = \omega\,\{\frac{-5}{8}\bar{Q}^4 A - \bar{Q}^2 B - \frac{1}{4}C - \frac{5}{32}\bar{Q}^4 - \frac{19}{24}\bar{Q}^2 - \frac{1}{12}\}$

$c_7 = \omega\,\{\frac{-35}{8}\bar{Q}^6 A - \frac{5}{2}\bar{Q}^4 B - \frac{1}{4}\bar{Q}^2 C - \frac{35}{32}\bar{Q}^6 - \frac{25}{24}\bar{Q}^4 - \frac{7}{12}\bar{Q}^2 + \frac{1}{3}\}$

$c_8 = \left(\frac{5}{8}\bar{Q}^2 + \frac{1}{2}\right)\bar{Q}^2 A + \left(\bar{Q}^2 + \frac{1}{2}\right)B + \frac{1}{4}C + \frac{5}{32}\bar{Q}^4 + \frac{11}{12}\bar{Q}^2 + \frac{5}{12}$

$c_9 = \left(\frac{35}{8}\bar{Q}^2 + \frac{15}{4}\right)\bar{Q}^4 A + \left(\frac{5}{2}\bar{Q}^2 + 3\right)\bar{Q}^2 B + \left(\frac{1}{4}\bar{Q}^2 + \frac{1}{2}\right)C + \frac{35}{32}\bar{Q}^6 + \frac{95}{48}\bar{Q}^4 + \frac{7}{3}\bar{Q}^2 + \frac{1}{6}$

$c_{10} = \left(\frac{35}{8}\bar{Q}^2 + \frac{15}{4}\right)\bar{Q}^4 A + \left(\frac{5}{2}\bar{Q}^2 + \frac{3}{2}\right)\bar{Q}^2 B + \frac{1}{4}\bar{Q}^2 C + \frac{35}{32}\bar{Q}^6 + \frac{95}{48}\bar{Q}^4 + \frac{5}{6}\bar{Q}^2 + \frac{1}{6}$

$c_{11} = \omega\,\{\left(\frac{-35}{8}\bar{Q}^2 - \frac{5}{2}\right)\bar{Q}^2 A - \left(\frac{5}{2}\bar{Q}^2 + 1\right)B - \frac{1}{4}C - \frac{35}{32}\bar{Q}^4 - \frac{5}{3}\bar{Q}^2 - \frac{3}{4}\}$

$c_{12} = \left(\frac{35}{8}\bar{Q}^4 + 5\bar{Q}^2 + 1\right)A + \left(\frac{5}{2}\bar{Q}^2 + 2\right)B + \frac{1}{4}C + \frac{35}{32}\bar{Q}^4 + \frac{55}{24}\bar{Q}^2 + \frac{7}{6}$

$c_{13} = \left(\frac{5}{8}\bar{Q}^2 + \frac{1}{2}\right)\bar{Q}^2 A - \frac{1}{2}\bar{Q}^2 B - \frac{1}{4}C + \frac{5}{32}\bar{Q}^4 - \frac{7}{12}\bar{Q}^2 - \frac{1}{12}$

$c_{14} = \omega\,\{\frac{-5}{8}\bar{Q}^4 A + \frac{1}{2}\bar{Q}^2 B + \frac{1}{4}C - \frac{5}{32}\bar{Q}^4 + \frac{17}{24}\bar{Q}^2 - \frac{1}{12}\}$

TABLE II. The structure of the conformally covariant gravitational polarization tensor $\tilde{\Pi}^{\mu\nu\alpha\beta}/\rho = \sum_{i=1}^{14} c_i T_i^{\mu\nu\alpha\beta}$ in terms of $A \equiv \tilde{\Pi}_{0000}/\rho$, $B \equiv \tilde{\Pi}_{0\mu}{}^\mu{}_0/\rho$, and $C \equiv \tilde{\Pi}_{\mu\nu}{}^{\mu\nu}/\rho$.